\documentclass[
referee
]{aastex61}

\usepackage{amsmath}
\usepackage{graphicx}
\usepackage{dcolumn}
\usepackage{bm}

\received{\today}
\revised{}
\accepted{}

\newcommand\aastex{AAS\TeX}

\submitjournal{ApJ}
\shorttitle{\aastex\ Lensing time delays with massive photons}
\shortauthors{Glicenstein}

\begin{document}


\title{Gravitational lensing time delays with massive photons}

\author{J-F. Glicenstein}
\affiliation{%
IRFU, CEA, UniversitŽ Paris-Saclay, F91191Gif-sur-Yvette, France\\
}%

\date{\today}

\begin{abstract}

This paper investigates the use of gravitational time delays of macro-lenses to constrain a possible photon mass. 
The time delay between the 2 compact images of a source of massive photons is computed. Explicit expressions are given for Schwarzschild and singular isothermal lenses. In the latter case, 
the time delay is very insensitive to   
the photon mass. Modeling lens galaxies  by a  singular isothermal model and a central supermassive black hole, the 
photon mass-dependent part of the time delay between the compact images is shown to be proportional to the mass of the black hole. The sensitivity of time delays to the photon mass is 
illustrated by a bound obtained from 3 AGN which have measurements in several passbands. The bound obtained is comparable to the limit with the deflection of radio waves by the Sun. 

\end{abstract}



\section {\label{sec:Introduction}Introduction} 
Gravitational deflection of massive particles has been studied by numerous authors, both in the general relativistic weak lensing limit \citep{1973PhRvD...8.2349L}, { strong lensing limit \citep{2014PhRvD..89h4075T}}and in semi-classical gravity \citep{2002CQGra..19.5429A,2004PhRvD..69j7501A}.
{ Massive neutral particles which play a role in astrophysics include neutrinos \citep{2001PhLB..512....8E, 2015JHEP...07..160C}. In the standard model of particle physics, photons are massless. They can acquire a mass without breaking gauge invariance by the Stueckelberg mechanism  in extensions of the standard model {\bf \citep{1938ActaHelvetica1,1938ActaHelvetica2,2004IJMPA..19.3265R}}. Another mass generating mechanism is the Higgs mechanism {\bf \citep{1964PhRvL..13..508H,1964PhRvL..13..321E,1964PhRvL..13..585G,2010RvMP...82..939G}}. The effective photon mass can depend on the spatial scale \citep{2007PhRvL..98a0402A}.} 
{ Most papers on lensing by massive particles concentrate on the angular deflection of light rays.}
Published results of the gravitational deflection of light rays by the Sun have been used to set limits on the photon mass, albeit not very competitive ones. { Recently, several authors have obtained photon mass limits from the astrometry of strong lensing systems \citep{2012SCPMA..55..523Q,2014MNRAS.437L..90E}. These papers do not use lens models and simply state that angular distances between images is equivalent to the deflection angle of light. This is a strong assumption which 
is discussed in appendix \ref{sec:astrometryappendix}.} The possibility to improve the photon mass lensing limits by 
using time delays from known macro-lenses is the subject of the present paper.  {Formulas for time delays of massive particles lensed by cosmic sources are not available in the literature to the best of the author's knowledge, or at least hard to find.}
The basic formulas for the time delay and position of images of a lensed massive photon source are consequently derived in the first section, assuming a weak gravitational field.  The photon-mass dependence of the time delay in several lensing models is discussed. The formulas obtained are then used to derive a bound on the photon mass from the time delays of AGN which have measurements in several passbands
Finally, strategies for improving the limit are discussed.   

\section{\label{sec:Formalism}Lensing of a massive photon source}
{ In this section, the basic formula for the time delay between lensed images of massive photon sources are derived.}
Massive photons are assumed to propagate in a spacetime described by the isotropic metric $g_{ij}$, given in cartesian coordinates by:
\begin{subequations}
\begin{eqnarray}
ds^2 &= g_{ij} dx^{i}dx^{j} = g_{00} dt^2 + g_{s s} d\vec{r}^2 \\
&= -(1+2U(r)) dt^2 + (1-2U(r))d\vec{r}^2 
\label{eq:1}
\end{eqnarray}
\end{subequations}
In equation (\ref{eq:1}), the speed of light c is set to 1 and the gravitational potential $U(r) \ll 1.$ In this paper, the metric signature is (-,+,+,+), greek letters are used for the spatial part of the metric
and $d\vec{r}^2 = dx^{\alpha} dx^{\beta} \delta_{\alpha \beta}$

The equations of motion of the massive particle in the background metric of equation \ref{eq:1} is described by the Hamiltonian:

\begin{equation}
H =  p_{0} = \sqrt{\frac{(g^{\alpha\beta}p_{\alpha}p_{\beta})+m^2}{-g^{00}}} = \sqrt{\frac{(g^{SS}\sum p_{\alpha}^2)+m^2}{-g^{00}}} .
\label{eq:massiveh}
\end{equation} 

\subsection{Travel time}\label{sec:traveltime}
The massive particle travel time from the source to the observer is evaluated in the weak gravity field, thin-lens and low-mass limits.
Hamilton's equations imply that $p_{0}$ is a constant of motion and that the velocity is 

\begin{equation}
\frac{dx^{\alpha}}{dt} =\frac{\partial H}{\partial p_{\alpha}} = \frac{g^{SS} p_{\alpha}}{-g^{00} p_{0}} \label{eq:Hamiltonx} 
 \end{equation}
 
 Using equations (\ref{eq:Hamiltonx}) and (\ref{eq:massiveh}) in the low mass, or high energy, limit ($\frac{m^2}{g^{00}p_{0}^2} \ll 1$)  gives the magnitude of the velocity
 \begin{equation}
 \frac{dr}{dt} = \sqrt{\frac{g^{ss}}{-g^{00}}} \sqrt{1 + \frac{m^2}{g^{00}p_{0}^2}} 
  \simeq \sqrt{\frac{g^{ss}}{-g^{00}}} \left( 1 + \frac{g_{00}}{2} \frac{m^2}{p_{0}^2} \right).
 \label{eq:drdt1} 
\end{equation}

Using equation (\ref{eq:drdt1}) and keeping only first order terms in  $\mu^2=\frac{m^2}{p_{0}^2}$ and $U(r),$ the expression for 
$\frac{dt}{dr}$ is readily obtained: 
\begin{subequations}
\begin{eqnarray}
 \frac{dt}{dr} &
  \simeq (1- 2U(r))\left( 1 + \frac{(1+2U(r))}{2}\mu^2 \right)  \\
 & \simeq  1 + \frac{1}{2} \mu^2 -2 U(r)
\end{eqnarray}
\end{subequations}

The travel time $T_{OS}$ is obtained by integrating $\frac{dt}{dr}$ over the massive photon trajectory from source to observer.
\begin{equation}
\label{eq:TOS}
    T_{OS} 
       = \int {dr \left( 1 + \frac{1}{2} \mu^2 -2 U(r) \right)} 
 \end{equation} 
 The structure of the $T_{OS}$ integral (\ref{eq:TOS}) reminds that of the "Fermat potential" encountered in (massless) photon lensing (\citet{1992grle.book.....S}, chapter 4). In the thin lens approximation, the massive particle moves 
 on a straight line until it gets deflected towards the observer. 
 The angular {\bf diameter} distance to the lens, located at redshift $z_{L}$ is $D_{OL},$ the angular {\bf diameter} distance from lens to observer is 
 $D_{LS}.$ Coordinates are taken in the plane perpendicular to the line of sight (the "lens plane"). The projected source position on the lens plane is at $\eta,$ and the impact parameter of the particle trajectory at $\zeta.$  
  
 Isolated stars or black holes can be modeled by Schwarzschild lenses. 
 The gravitational potential of a Schwarzschild lens of mass $M_{L}$ is $U(r) = -\frac{GM_{L}}{r}.$ Integration of equation (\ref{eq:TOS}) yields  
\begin{widetext}
 \begin{equation}
\label{eq:TimeSchwarzschild}
    T_{OS} =  (z_{L}+1) \left(\frac{1}{2} (1 +\frac{1}{2} \mu^2) (\frac{1}{D_{OL}} + \frac{1}{D_{LS}})(\zeta-\eta)^2-4GM_{L} \ln{\zeta} +T_0\right) ,
 \end{equation} 
 \end{widetext}
where $T_0$ is a constant. The travel time $T_{OS}$ is multiplied by a factor of $(z_L+1)$ to take the redshift of the lens into account.  

Galaxy lenses, which produce macro-lenses, are not well described by Schwarzschild lenses, but rather by extended mass models. A popular model is the 
Singular Isothermal Lens (SIL) model and its elliptical extensions. In the SIL model, 
$U(r) = 2\sigma_v^2 \ln{r} $ where $\sigma_v$ is the velocity dispersion of stars in the lens galaxy.  The travel time for source to observer in a SIL model is: 
\begin{widetext}
 \begin{equation}
\label{eq:TimeSIL}
    T_{OS}  = (z_{L}+1) \left( \frac{1}{2} (1 +\frac{1}{2} \mu^2) (\frac{1}{D_{OL}} + \frac{1}{D_{LS}})(\zeta-\eta)^2-2\pi\sigma_v^2|\zeta|+T_0 \right)
 \end{equation}
  \end{widetext}

Since equation (\ref{eq:TOS}) is linear in $U(r),$ the travel time of a photon lensed by a galaxy with a central supermassive black hole (SMBH) is the sum of the contributions of the galaxy (equation (\ref{eq:TimeSIL})) and 
the black hole (equation (\ref{eq:TimeSchwarzschild})).
  
In massless photon lensing, the lens equation, giving the position of images $\zeta$ as a function of the source position $\eta,$ is obtained by differentiating $T_{OS}$ with respect to $\zeta,$ since the travel time 
is extremal on the photon path. This property is no longer valid for massive photons.  In the next section, the lens equation is obtained directly from the deflection angle.

\subsection{Lens equation and image positions}

The line of sight to the source is taken as the $z$ direction. Since the deflection is small, $p_z \gg p_x,  p_y. $ Setting $P^2 = p^{\alpha}p^{\beta} \delta_{\alpha \beta},$ Hamilton's equation for $p_{x}$ gives

\begin{subequations}
\begin{eqnarray}
\frac{dp_{x}}{dt} &= -\frac{\partial H}{\partial x} 
                           = -\partial_{x} \left( ( \sqrt{\frac{g^{SS}P^2}{-g^{00}}})(1+\frac{1}{2}  \frac{m^2}{g^{SS}P^2}) \right)  \\
                          & = -P\left(\partial_{x}\sqrt{\frac{g^{SS}}{-g^{00}}} + \frac{1}{2}  \partial_{x}(\frac{m^2}{P^2})      \right) , 
                          \label{eq:dpdx}
\end{eqnarray}
\end{subequations}
where equation (\ref{eq:dpdx}) uses the equality $-g^{00}g^{SS} = 1 + O(U(r)^2).$  Since $\sqrt{\frac{g^{SS}}{-g^{00}}} \simeq g^{SS} \simeq 1+2U(r),$ the change in $x$ momentum is
\begin{equation}
\frac{dp_{x}}{dt}  \simeq  -2P \partial_{x} U(r)
\label{eq:deflection}
\end{equation}.

In equation (\ref{eq:deflection}), $\frac{dp_{x}}{dt}$ is a first order quantity in $U(r).$ The calculation of the deflection angle uses the integration of 
$\frac{dp_{x}}{dz}$ over $z.$ It is thus only necessary to evaluate $dz/dt$ to zeroth order in $U(r).$  Setting $g^{00}=g^{SS} = 1$ in equation 
(\ref{eq:Hamiltonx}) gives 
\begin{equation}
\frac{dz}{dt} = \frac{p_z}{p_0} + O(U(r)) 
\simeq 1 - \frac{1}{2}  \mu^2 + O(U(r))   
\end{equation}

The deflection angle $\alpha$ is obtained by integrating the relative momentum change  $\frac{1}{P} \frac{dp_{x}}{dz}$ over $z$
\begin{subequations}
\begin{eqnarray}
\alpha &= \int{dz \frac{1}{P} \frac{dp_{x}}{dz}} = -\int{dz\frac{ \partial_{x} U(r)}{ 1 - \frac{1}{2}  \mu^2 }} \\
&= -\int{dz(1 + \frac{1}{2} \mu^2)\partial_{x} U(r)} \label{eq:alpha}
\end{eqnarray}
\end{subequations}

 Inspection of equation (\ref{eq:alpha}) shows that the deflection angle is as for massless photons, except for the $(1 + \frac{1}{2} \mu^2)$ multiplicative factor.
 For a Schwarzschild lens and a massive photon with impact parameter $\zeta$, the deflection angle is
\begin{equation}
 \alpha = \frac{4GM_L}{\zeta} (1 + \frac{1}{2}\mu^2). 
\label{eq:alphaSch}
 \end{equation}
 This result was first derived by \citet{1973PhRvD...8.2349L}. 

For a SIL lens, the absolute value of the deflection angle is independent of $|\zeta|:$
\begin{equation}
 \alpha = 2\pi\sigma_v^2 (1 + \frac{1}{2}\mu^2)\mbox{sgn}(\zeta). 
\label{eq:alphaSIL}
 \end{equation}

The deflection angle of a lens composed of a galaxy (modeled by a SIL) hosting a SMBH in its centre is the sum of the deflection angles of the components.

The lens equation is derived by expressing the deflection angle as
\begin{equation}
\alpha(\zeta) =(\frac{1}{D_{OL}} + \frac{1}{D_{LS}})(\zeta-\eta) 
\end{equation}

For a Schwarzschild lens, the lens equation has the form
\begin{equation}
\zeta-\eta = \frac{r_{E}^2}{\zeta},
\label{eq:PLPS}
\end{equation}
with an energy dependent Einstein radius
  \begin{equation}
\label{eq:EinsteinRadius}
  r_{E}^2(\mu^2) = \frac{4GM_{L} (1 + \frac{1}{2}\mu^2)D_{OL}D_{LS}}{D_{OS} }.
 \end{equation}

For a SIL lens, the lens equation is:
 \begin{equation}
\zeta-\eta = \mbox{sgn}(\zeta) l_{E},
\label{eq:SILPS}
\end{equation}
with $l_{E}$ defined by
  \begin{equation}
\label{eq:EinsteinLength}
  l_{E}(\mu^2) = \frac{4\pi \sigma_v^2 (1 + \frac{1}{2}\mu^2) D_{OL}D_{LS} }{(D_{OL}+D_{LS})}
 \end{equation}

\section{Mass dependence of the time delay}\label{sec:massedepenceoftd}
Using equation (\ref{eq:term1}) and (\ref{eq:term2})  and working to first order in $\mu^2,$ one finds the mass dependent 
part of the time delay between the images of a Schwarzschild lens
 \begin{equation}
\label{eq:TimeSchwarzschild}
    \Delta T_{\mu^2} =  - \frac{(z_{L}+1)\mu^2}{4}  (\frac{1}{D_{OL}} + \frac{1}{D_{LS}})\eta \sqrt{\eta^2+4r_E^{2}(0)} .
 \end{equation}
 For the SIL model, the time delay between the images is independent of the photon mass, as shown in the appendix. The independence of the time delay on the 
 photon mass is not true for general singular isothermal elliptical models. The analysis of the ellipticity induced time delay 
 is outside the scope of this paper. 

Since the time delay between compact images is independent of the photon mass for the simplest galactic model, black holes and especially SMBH provide much cleaner environments than galaxies to constrain the photon mass
with lensing time delays. However, except possibly for the Galactic Center black hole, lensing by SMBH and by its host galaxy cannot be disentangled. The influence of the central SMBH on time delay can be conveniently studied
with a model having a Schwarzschild lens at the center of a SIL, as in  \citet{2012MNRAS.420..792M}.   

The travel time for a massive particle lensed  by the sum of a Schwarzschild and an isothermal galaxy with the same center is:
\begin{equation}
    T_{OS}  =(z_{L}+1)  (\frac{1}{D_{OL}} + \frac{1}{D_{LS}}) \left( \frac{1}{2}(1 + \frac{1}{2}\mu^2) (\zeta -\eta)^2-l_{E}(0) |\zeta| -r_{E}^2(0)\ln{|\zeta|} \right)
  \label{eq:Fermat16}   
 \end{equation} 
The lens equation is obtained from the deflection angles:
\begin{equation}
   \zeta-\eta-l_{E}(\mu^2) \mbox{sgn}(\zeta) -\frac{r_{E}^2(\mu^2)}{\zeta}=0
  \label{eq:Fermat17}   
 \end{equation} 
 It has  2 solutions $\zeta_{\pm}$ defined by
 \begin{subequations}
 \begin{eqnarray}
 \zeta_{+} &= \frac{1}{2}(\eta + l_{E}(\mu^2) + \sqrt{(\eta+l_E(\mu^2))^2 + 4r_E^2(\mu^2)}) \label{eq:solcombined} \\
 \zeta_{-} &= \frac{1}{2}(\eta-l_{E}(\mu^2) - \sqrt{(\eta-l_E(\mu^2))^2 + 4r_E^2(\mu^2)}) 
 \label{eq:combinedsol}
 \end{eqnarray}
 \end{subequations}

The mass of the central black hole is correlated to the central velocity dispersion of its host galaxy through the M-$\sigma$ relation \citep{2000ApJ...539L...9F,2000ApJ...539L..13G}. Since the 3 AGNs analyzed in this paper
are lensed by late-type galaxies, the mass of the central black hole is estimated by the relation \citep{2009ApJ...698..198G}.
\begin{equation}
\frac{M_{L}}{M_{\odot}} = 10^{7.95} \left( \frac{\sigma_c}{200 \mbox{km/s}} \right)^{4.58} 
\end{equation} 
ignoring the scatter in the exponents. $\sigma_c$ is the central velocity dispersion and is equal to the previously defined $\sigma_v$ parameter in the SIL model. 
Assuming $(\frac{D_{OL}D_{LS}}{D_{OS}})\simeq 0.5$ Gpc and $\sigma \simeq 200$ km/s, the ratio $ \frac{ r_{E}^2}{l_{E}^2}$ is of the order of 1\%. 

It is shown in the appendix that the time delay between the images, assuming $\eta \leq l_{E}$ is
\begin{equation}
    \Delta T  =(z_{L}+1)  (\frac{1}{D_{OL}} + \frac{1}{D_{LS}}) \left( -2l_{E}(0)\eta - r_{E}^2(0) \ln{(\frac{(\eta + l_{E})(0)}{(-\eta + l_{E}(0)) })} + \mu^2 \eta l_{E}(0) \frac{r^2_{E}(0)} {\eta^2-{l}_E^2(0)} \right)    \label{eq:ISOSchwfinal}   
 \end{equation} 
to first order in $\frac{r^2_{E}}{l_{E}^2}$ and $\mu^2.$ 

The mass-dependent part of the time delay $\Delta T$ is thus proportional to the mass of the SMBH of the lensing galaxy. 
{ In the limit $\eta \ll l_{E},$    equation (\ref{eq:ISOSchwfinal}) simplifies further to
\begin{equation}
    \Delta T  \simeq \Delta T(0) (1+ \frac{1}{2} \frac{r^2_{E}}{l_{E}^2} \mu^2) \simeq  \Delta T(0) (1+ \frac{1}{2} 10^{-2} \mu^2)    \label{eq:ISOSchwsimplified}   
 \end{equation} 
}

\section{Bounds on the photon mass from gravitational lensing}\label{sec:bounds}

Only 3 gravitational lensed AGN with only 2 compact images have time delays measured in the radio passband in the sample of \citet{2015A&A...580A..38R}, namely JVAS B0218+357, CLASS B1600+434 and
PKS 1830-211.  It turns out that these AGN have also measurements of their time delays in other passbands. These measurements are listed in Table \ref{tab:measures}.
\begin{deluxetable}{cccc}
\tablecaption{Measurements of time delays in JVAS B0218+357, CLASS B1600+434 and PKS 1830-211 \label{tab:measures}}
\tablehead{
\colhead{AGN} &
\colhead{Passband} &
\colhead{Time delay (day)} &
\colhead{Reference}}
\startdata
JVAS B0218+357 & 8 GHz & $9.6^{+1.3}_{-1.2}$ (95\% CL) & \citet{2000ApJ...545..578C} \\
& 8.4 GHz & $10.1^{+1.4}_{-0.7}$ (95\% CL) &  \citet{1999MNRAS.304..349B} \\
& 15 GHz &  $11.3^{+2}_{-1.8}$ (95\% CL) & \citet{2000ApJ...545..578C}  \\
& 15 GHz &  $10.6^{+0.7}_{-0.5}$ (95\% CL)  & \citet{1999MNRAS.304..349B} \\
& 0.1-30 GeV  & $11.46 \pm 0.16$ (68\% CL) & \citet{2014ApJ...782L..14C} \\
CLASS B1600+434  & 8 GHz & $47^{+5}_{-6}$ (68\% CL) & \citet{2000AA...356..391K} \\
 & I band & $47.8\pm 1.2$ (68\% CL) & \citet{2000ApJ...544..117B, 2011AA...536A..44E}\\
 PKS 1830-211  & 8.6 GHz & $26^{+5}_{-4}$ (95\% CL) & \citet{1998ApJ...508L..51L} \\
 & millimiter band & $24^{+5}_{-4}$ (95\% CL) & \citet{2001ASPC..237..155W}  \\
 & 0.3-30 GeV & $27.1\pm0.6$ (68 \% CL) & \citet{2011AA...528L...3B}  \\
\enddata
\end{deluxetable}

%


Bounds on the photon mass are obtained by looking for the energy-dependent time delay given by equation (\ref{eq:ISOSchwfinal}). { 
Several systematic effects in the modeling of lenses and in the measurement of time delays have to be considered.
The measured time delay is affected by the location of emission in the AGN and by plasma dispersion effects.
The angular resolution of radio telescopes is sufficient to identify the origin, core or jet, of the radio emission of the lensed AGN. 
 In contrast, the Fermi-LAT instrument, which provided the measurements of  \citet{2014ApJ...782L..14C}  and \citet{2011AA...528L...3B}, does not have a sufficient angular resolution to resolve the various components of the lensed AGN.
 Because of this, the high energy time delays reported in Table \ref{tab:measures}  may not have the same spatial origin as the radio time delays.  

Light dispersion by the galactic plasma around lenses  is also a possible source of systematic effects.  Photons propagate in a plasma with an effective mass equal to the plasma frequency 
$\omega_p \simeq 3.7\ 10^{-11} \left( n_{e}/\mathrm{cm}^{-3} \right)^{1/2}\ \mathrm{eV},$ where $n_{e}$ is the electron density \citep{2015PlPhR..41..562B}. 
The time delay formulas from the present paper are also valid for photon propagation in a uniform plasma with equivalent mass $\omega_p.$  The typical value of Galactic electron 
densities is $n_e \simeq 0.01\mathrm{cm}^{-3},$ with fluctuations of up to 2 orders of magnitude \citep{2017ApJ...835...29Y}.
Taking $10\ \mathrm{cm}^{-3}$ as the maximum value of the electron density in the lensing galaxies, the equivalent photon mass in the galactic plasma is less than $10^{-10}  \mathrm{eV}.$  This is several orders of magnitude less than the limit on the photon mass
obtained in this paper, so that light dispersion by galactic plasmas can be safely neglected.

Lens modeling uncertainties were already mentioned in section \ref{sec:massedepenceoftd}.} The ellipticity of the lensing galaxy is neglected and could give an independent, contribution
to the mass-dependent part of the time delay.  
The contribution of the SMBH tends to increase the mass-dependent part of the time delay at low photon energies, according to equation   (\ref{eq:ISOSchwfinal}).
The ellipticity of the lens galaxy and the influence of neighboring masses, producing shear, may increase or decrease it. 

The lenses analyzed in this paper were not modeled in detail and time delays were fitted to the simplified formula (\ref{eq:ISOSchwsimplified}). { The values $ \Delta T(0)$ and $\mu^2$ are first fitted independently for each source, using a fitting program from the ROOT library \citep{1997NIMPA.389...81B}. The fitted values are reported in Table \ref{tab:results}.
\begin{deluxetable}{ccc}
\tablecaption{Fitted time delays and reduced mass squared $\mu^2$ in JVAS B0218+357, CLASS B1600+434 and PKS 1830-211 \label{tab:results}}
\tablehead{
\colhead{AGN} &
\colhead{Fitted time delay (day)} &
\colhead{$\mu^2$ (eV$^2$)}}
\startdata
JVAS B0218+357 &  $11.4\pm0.15$ & $(-2.9\pm0.6)\ 10^{-8}$ \\
CLASS B1600+434  & $47.8\pm1.7$ & $(-0.4\pm4.1)\ 10^{-8}$ \\
 PKS 1830-211  & $26.9\pm0.7$ & $(-0.8\pm1.9)\ 10^{-8}$ \\
\enddata
\end{deluxetable}

Next, the time delays of the 3 AGN are normalized to their fitted $\Delta T(0)$ and the combined data are used to refit the value of $\mu^2.$
The combined fitted value is
${\mu^2}^{fit} = -2.6\ 10^{-8} \pm 6.8\ 10^{-9} \mbox{eV}^2.$   
 The negative measurement of $\mu^2$ reflects systematic errors, especially the different AGN emission regions at radio and high energies. A conservative limit can be  obtained by relying solely  on the error bar of $\mu^2,$ which  
 measures the sensitivity of the analysis to a possible photon mass. Assuming a null result in the mass dependence of the time delays, the 90\% CL upper limit  is thus $m_{\gamma} \leq 9.3\ 10^{-5}$ eV.  This limit is obtained at the galactic
 (10 kpc) spatial scale.
}


\section{Conclusion}

The bound on the photon mass derived in section \ref{sec:bounds} is several orders of magnitude less constraining than the best limits available \citep{2005RPPh...68...77T}, 
but comparable to the limit obtained with the gravitational deflection of radio waves by the Sun \citep{2010PhRvD..82f5026A}. 
There are several ways to improve the result of section \ref{sec:bounds}. The lens could be modeled with more realistic models, taking into account the ellipticity of the mass distribution and the external shear. Several four-image lensing systems, such as Q0957+561, have been  monitored in several passbands and could then be included in the analysis with a proper lens modeling. 
{ Combining with astrometry, as in the approach of \citet{2012SCPMA..55..523Q} and \citet{2014MNRAS.437L..90E} will further improve the constraints.} 
Including a  model of the source emission \citep{2016ApJ...821...58B}  would reduce the systematic uncertainties on the time delay measurement. Finally, the number of observed strong lensing systems will increase drastically in the near future. 
The Euclid mission \citep{2013LRR....16....6A} and the LSST \citep{2009arXiv0912.0201L} are likely to monitor $10^{3}$  strong lensing systems with measurable time delays in the optical and IR passbands. SKA, operating at 1.4 GHz, will discover $\sim 10^5$ lensing systems and will measure hundreds of time delays \citep{2004NewAR..48.1085K}.  Combining the improvement in statistics and systematics, the bound on the photon mass from gravitational macrolensing could be lowered by one or two orders of magnitude.

\appendix
\section{Mass dependence of the solutions of the lens equation}\label{sec:massdependence}
\subsection{Schwarzschild lens}
Equation (\ref{eq:PLPS}) has 2 solutions given by 
\begin{equation}
\zeta_{\pm} = \frac{1}{2}(\eta\pm \sqrt{\eta^2+4r_E^2})
\end{equation}
Since $\mu^2$ is assumed small compared to 1, the solutions can be expanded as
\begin{subequations}
\begin{eqnarray} 
\zeta_{\pm} &= \zeta_{\pm}(0) \pm \frac{1}{\sqrt{\eta^2+4r_E^{2}(0)}} \frac{\partial r_{E}^2}{\partial \mu^2} \mu^2 \\
& = \zeta_{\pm}(0) \pm \frac{r_{E}^2(0)}{2\sqrt{\eta^2+4r_E^{2}(0)}} \mu^2
\label{eq:zetapm}
\end{eqnarray}
\end{subequations}
where $r_{E}(0), \zeta_{\pm}(0) $ are short-hand notations for $ r_{E}(\mu^2=0)$ and $\zeta_{\pm}(\mu^2=0).$

The calculation of time delay uses 2 combinations of $\zeta_{\pm},$ namely $\ln{\frac{\zeta_{-}}{\zeta_{+}}}$ and $\zeta_{-}^2-\zeta_{+}^2.$
Equation (\ref{eq:zetapm}) shows that 
\begin{subequations}
\begin{eqnarray}
\ln{\left(\frac{\zeta_{-}}{\zeta_{+}}\right)} 
 &\simeq \ln{\left(\frac{\zeta_{-}(0)}{\zeta_{+}(0)}\right)}  
 -\frac{r_{E}^2(0)}{2\sqrt{\eta^2+4r_E^{2}(0)}} \mu^2 \left( \frac{1}{\zeta_{-} (0)}+  \frac{1}{\zeta_{+}(0)} \right) \\
& = \ln{\frac{\zeta_{-}(0)}{\zeta_{+}(0)}} +\frac{\eta}{2\sqrt{\eta^2+4r_E^{2}(0)}} \mu^2.
\label{eq:term1}
\end{eqnarray}
\end{subequations}

The other combination is:
\begin{subequations}
\begin{eqnarray}
\zeta_{-}^2-\zeta_{+}^2 &= -\eta \sqrt{\eta^2+4r_E^{2}(\mu^2)} \\
&= -\eta \sqrt{\eta^2+4r_E^{2}(0)}\left( 1+ \frac{r_E^2(0)}{\eta^2+4r_E^2(0)} \mu^2 \right)
\label{eq:term2}
\end{eqnarray}
\end{subequations}

\subsection{SIL model}
Equation (\ref{eq:SILPS}) has 2 solutions for $|\eta| \le l_{E}.$ These solutions are
\begin{eqnarray}
\zeta_{+} & = l_{E} + \eta \label{eq:combinedsol2.5}  \\
\zeta_{-} & = \eta - l_{E} \label{eq:combinedsol2.6}
\end{eqnarray}
Since $(\zeta_{+}-\eta)^2 = (\zeta_{-}-\eta)^2 = l_{E}^2,$  only the second term in equation  
(\ref{eq:TimeSIL}) contributes to the time delay between images. This term depends on 
\begin{equation}
|\zeta_{+}| - |\zeta_{-}| = 2\eta.
\end{equation}
 It is independent of $l_{E},$ so that the time delay between images in the SIL model is independent of $\mu^2.$

\subsection{SIL model with a central SMBH}

Since $\frac{r_{E}}{l_{E}} \sim 0.1$ for the AGNs under study, the solutions of the lens equation (\ref{eq:Fermat17}) are first developed to first order in $\frac{r^2_{E}}{l_{E}^2}.$
\begin{subequations}
\begin{eqnarray}
 \zeta_{+} &\simeq \frac{1}{2}(\eta + l_{E} + (\eta + l_{E})(1+ \frac{2r_E^2}{(\eta + l_{E})^2})) \\
 & = (\eta + l_{E}) + \frac{r_E^2}{(\eta + l_{E})} \label{eq:combinedsol3.5} \\
 \zeta_{-} &\simeq \frac{1}{2}(\eta-l_{E} - |(\eta-l_E)|(1 + \frac{2r_E^2}{(\eta - l_{E})^2} )) \\
 & = (\eta - l_{E}) + \frac{r_E^2}{(\eta - l_{E})}
 \label{eq:combinedsol4}
 \end{eqnarray}
 \end{subequations}
 
 The components of the time delay equation  (\ref{eq:Fermat16}) are expanded in powers of $r_{E}^2,$ keeping only the lowest order.
\begin{subequations}
\begin{eqnarray}
 (\zeta_{+} -\eta)^2 -  (\zeta_{-} -\eta)^2  &= 2l_{E}r_E^2 (\frac{1}{(\eta + l_{E})} + \frac{1}{(\eta - l_{E})}) 
 = \frac{4\eta l_{E}r_E^{2}}{\eta^2-l_E^2}  \\
 |\zeta_{+}| -|\zeta_{-}|
 &= 2\eta + r_E^2 (\frac{1}{(\eta + l_{E})} + \frac{1}{(\eta - l_{E})}) 
= 2\eta + \frac{2\eta r_E^2}{\eta^2-l_E^2} \\
r_{E}^2(0) \ln{(\frac{|\zeta_{+}|}{|\zeta_{-}|})} 
 & = r_{E}^2(0) \ln{(\frac{(\eta + l_{E})}{(-\eta + l_{E}) })} +O(r_{E}^4) 
 \label{eq:combinedsol7}
 \end{eqnarray}
 \end{subequations}
 
 Summing up the terms, equation   (\ref{eq:Fermat16}) changes to
\begin{equation}
    \Delta T  =(z_{L}+1)  (\frac{1}{D_{OL}} + \frac{1}{D_{LS}}) \left( (1+\frac{\mu^2}{2})\frac{2\eta l_{E}(\mu^2){r}_E^{2}(\mu^2)}{\eta^2-{l}_E^2(\mu^2)} -l_{E}(0)(2\eta + \frac{2\eta{r}_E^2(\mu^2)}{\eta^2-{l}_E^2(\mu^2)} )  - r_{E}^2(0) \ln{(\frac{(\eta + l_{E}(\mu^2))}{(-\eta + l_{E}(\mu^2) ) })} \right)
  \label{eq:Fermat26}   
 \end{equation}
 
The next step is to expand the expression of $\Delta T$ from equation (\ref{eq:Fermat26}) as a function of $\mu^2.$
The 2 first terms in the bracket on the right hand side
can be summed together to obtain  
\begin{subequations}
\begin{eqnarray}
    (1+\frac{\mu^2}{2})\frac{2\eta l_{E}(\mu^2){r}_E^{2}(\mu^2)}{\eta^2-{l}_E^2(\mu^2)} -l_{E}(0)(2\eta + \frac{2\eta{r}_E^2(\mu^2)}{\eta^2-{l}_E^2(\mu^2)} ) &= -2l_{E}(0)\eta + 2\eta l_{E}(0) \frac{r^2_{E}(\mu^2)( (1+\frac{\mu^2}{2})^2-1)} {\eta^2-{l}_E^2(\mu^2)} \\
   & = -2l_{E}(0)\eta + 2 \mu^2 \eta l_{E}(0) \frac{r^2_{E}(0)} {\eta^2-{l}_E^2(0)} + O(\mu^4)
\label{eq:Fermat28}
 \end{eqnarray} 
 \end{subequations}

The third term in the bracket is:
\begin{equation}
-r_{E}^2(0)\ln{(\frac{(\eta + l_{E}(\mu^2))}{(-\eta + l_{E}(\mu^2)) })} 
 = -r_{E}^2(0)\ln{(\frac{(\eta + l_{E}(0))}{(-\eta + l_{E}(0)) })}-\frac{\mu^2 r_{E}^2(0)\eta l_E(0)}{\eta^2 -l_E^2(0)}
\label{eq:Fermat27}
\end{equation}
Finally, the time delay (equation (\ref{eq:ISOSchwfinal})) is obtained by adding the results of equations (\ref{eq:Fermat28}) and (\ref{eq:Fermat27}). 
 
{ \section{Dependence of the distance between images with photon mass}\label{sec:astrometryappendix}
Several authors \citep{2012SCPMA..55..523Q,2014MNRAS.437L..90E} bound the photon mass with a multiwavelength astrometric study of strong lensing systems. They further assume that 
\begin{equation}
\label{eq:astrometricclaim}
{\zeta_{+}({\mu}^2) -\zeta_{-}({\mu}^2)}=  (1+k\mu^2)({\zeta_{+}(0) -\zeta_{-}(0)} )
\end{equation}
where $k$ is a coefficient of order unity and independent of the lens model. This appendix discusses the validity of this assumption.

The validity of the claim is first examined with the 2-images models used in the present work. 
The mass dependent part of the distance between images in the Schwarzschild and SIL models can be obtained from equations
(\ref{eq:zetapm}) and (\ref{eq:combinedsol2.5}-\ref{eq:combinedsol2.6}). 
The mass dependent part of the distance between the images in the  {\bf circular} SIL model is
\begin{equation}
\frac{\zeta_{+}({\mu}^2) -\zeta_{-}({\mu}^2)}{\zeta_{+}(0) -\zeta_{-}(0)} =  1+\frac{1}{2}\mu^2
\end{equation}
which is equivalent to light deflection,  {\bf in support of the claims of  \citet{2014MNRAS.437L..90E}. 
Unfortunately, this property of lenses does not hold for extensions of the circular SIL model, as shown by equation (\ref{eq:b16}), derived for a circular SIL model with a central black hole}.

In the case of the Schwarzschild model, the mass dependent part of the distance between images is
\begin{equation}
\frac{\zeta_{+}(\mu^2) -\zeta_{-}(\mu^2)}{\zeta_{+}(0) -\zeta_{-}(0)} =  1+\frac{1}{2}\left(\frac{r_{E}^2}{\eta^2+4r_{E}^2} \right)\mu^2
\end{equation}
The coefficient of $\frac{1}{2}\mu^2$ depends on the source position and varies {\bf formally} from 0 to 1/4. {\bf The actual lower limit depends on the sensitivity of the observing instrument.} 
The perturbation of the SIL model with a central SMBH, keeping the circular symmetry, can be studied directly with equations 
 (\ref{eq:solcombined}-\ref{eq:combinedsol}). The mass dependent part of the distance between the images in this model is a complicated function of $\zeta,$ $l_{E}$ and $r_{E}.$ However, the asymptotic behavior
 of  $\frac{\zeta_{+}(\mu^2) -\zeta_{-}(\mu^2)}{\zeta_{+}(0) -\zeta_{-}(0)}$ when the source position $\zeta$ goes to infinity is readily obtained.
 \begin{equation}
 \label{eq:b16}
\frac{\zeta_{+}(\mu^2) -\zeta_{-}(\mu^2)}{\zeta_{+}(0) -\zeta_{-}(0)} \sim 1+\frac{1}{2}\frac{l_{E}}{\lvert \eta \rvert } \mu^2
\end{equation}
In the SIL model with a central SMBH, the coefficient of $\frac{1}{2}\mu^2$ can also be {\bf formally} arbitrarily small, {\bf limited by the sensitivity of the observing instrument}.  
To obtain a limit on $\mu^2,$ the $k$ coefficient from equation (\ref{eq:astrometricclaim}) has to be bounded from below. Obviously, this is not possible in a model independent way. The AGN source position has to be taken 
into account.

The source position is expected to have a stronger impact in multi-image lensing systems. In these systems, pair of images can appear or disappear depending on the source position relative to caustics.  The position of the caustics itself depends on $\mu^2,$ as seen in the SIL model.

}


\bibliography{massivephotonlensing}

\end{document}